\documentclass[letterpaper,10pt,conference]{ieeeconf}
\IEEEoverridecommandlockouts
\usepackage[english]{babel}

\usepackage{amsmath,amssymb,latexsym,amsfonts,mathptmx}
\usepackage{xcolor,graphicx,subfigure}
\usepackage{enumerate}
\usepackage{epsfig}
\usepackage[lined,linesnumbered,algoruled]{algorithm2e}
\catcode`\<=\active \def<{
        \fontencoding{T1}\selectfont\symbol{60}\fontencoding{\encodingdefault}}
\newcommand{\nin}{\not\in}

\newcommand{\tmtextit}[1]{{\itshape{#1}}}

\newtheorem{theorem}{Theorem}[section]

\newtheorem{assumption}[theorem]{Assumption}

\newtheorem{lemma}[theorem]{Lemma}

\newtheorem{proposition}[theorem]{Proposition}

\newcommand{\oprocendsymbol}{\hbox{$\bullet$}}
\newcommand{\oprocend}{\relax\ifmmode\else\unskip\hfill\fi\oprocendsymbol}

\newcommand{\real}{{\mathbb{R}}}
\newcommand{\naturals}{{\mathbb{N}}}

\renewcommand{\theenumi}{(\roman{enumi})}

\newcommand{\II}[1]{\mathcal{I}^{#1}}
\newcommand{\zeros}{\bold{0}}
\newcommand{\ones}{\bold{1}}

\newcommand{\longthmtitle}[1]{\tmtextit{(#1).}}
\newcommand{\myclearpage}{\clearpage}
\renewcommand{\myclearpage}{}

\usepackage[normalem]{ulem}

\renewcommand{\baselinestretch}{.98}

\allowdisplaybreaks

\myclearpage
\begin{document}
        
\title{\Large \bf Transient frequency control with regional
  cooperation for power networks\thanks{This work is supported by
    AFOSR Award FA9550-15-1-0108.}}

\author{Yifu Zhang and Jorge Cort{\'e}s\thanks{The authors are with
    the Department of Mechanical and Aerospace Engineering, University
    of California, San Diego, CA 92093, USA, {\tt\small
      \{yifuzhang,cortes\}@ucsd.edu}}}
\maketitle

\begin{abstract}
  This paper proposes a centralized and a distributed sub-optimal
  control strategy to maintain in safe regions the real-time transient
  frequencies of a given collection of buses, and simultaneously
  preserve asymptotic stability of the entire network. In a receding
  horizon fashion, the centralized control input is obtained by
  iteratively solving an open-loop optimization aiming to minimize the
  aggregate control effort over controllers regulated on individual
  buses with transient frequency and stability constraints. Due to the
  non-convexity of the optimization, we propose a convexification
  technique by identifying a reference control input trajectory. We
  then extend the centralized control to a distributed scheme, where
  each sub-controller can only access the state information within a
  local region. Simulations on a IEEE-39 network illustrate our
  results.
\end{abstract}


\section{Introduction}\label{section:intro}

Power network transient stability refers to the ability of an electric
power network to remain synchronized after disturbances, during which
system states should stay within safe bounds so that the entire system
remains physically intact~\cite{PK-JP:04}. Due to the dynamics and
interconnection of power networks, even if the power supply and
consumption are re-balanced immediately after a failure, individual
generators are still in the danger of overheating due to large
transient frequency or voltage deviations, which in turn may trigger
cascading failures. In practice, it is also common to treat the
transient frequencies of some crucial generators as a key metric for
evaluating system performance or as indexes for applying load-shedding
strategies~\cite{NWM-KC-MS:11}.
These considerations motivate us to design a frequency controller that
mitigates the frequency overshoot observed in transients, and at the
same time, preserves synchronization of the whole system.
        
\textit{Literature review:} A body of work~\cite{HDC:11,FD-MC-FB:13}
studies how network synchronization relates to factors such as network
topology, parameter values, initial conditions, and power
supply-demand balance. However, there is no guarantee that transient
frequencies of individual buses do not exceed their physical limits,
and thus, synchronization may not necessarily hold under transient
frequency constraints. To improve transient behavior, various
strategies have been proposed including power
re-dispatch~\cite{AA-EBM:06}, power system stabilizer
(PSS)~\cite{PK:94}, and virtual inertial
placement~\cite{TSB-TL-DJH:15}. Nonetheless, these strategies do not
rigorously ensure that transient state stay within a safe region. Our
previous work combines Lyapunov stability analysis and barrier
function to propose a distributed controller~\cite{YZ-JC:18-cdc1} that
simultaneously guarantees both synchronization and transient frequency
safety.  On the other hand, to account for the trade-offs between
performance and control effort,~\cite{ANV-IAH-JBR-SJW:08,DJ-BK:02}
investigate (distributed) model predictive control~(MPC) for networked
system. The work~\cite{DJ-BK:02} treats each subsystem as an
independent system by considering the effect of interconnections as
bounded uncertainties, resulting in a conservative approach to
establishing stability. The work~\cite{ANV-IAH-JBR-SJW:08} shows that
each subsystem having no knowledge of others' cost
functions~\cite{EC-DJ-BHK-ST:02} leads to a non-cooperative game, and
the control input trajectory may even diverge.  In addition, to
maintain the distributed nature of MPC, the predicted horizon is
limited to a single step~\cite{ANV-IAH-JBR-SJW:08,MHN-etal:14} to
restrict information sharing. As the horizon increases, the
distributed control could require global information.

\textit{Statement of contribution:}
We develop a distributed MPC framework that meets the following
requirements on system performance, control cost, and control
structure: (i)~all bus frequencies converge to the same (potentially
unknown) frequency; (ii)~for each bus of interest, if its initial
frequency belongs to a desired safe frequency region, then its
frequency trajectory stays in it for all subsequent time; (iii)~for
each region of the network, sub-controllers within it cooperatively
achieve requirements~(i) and~(ii) by reducing their overall control
efforts; and (iv)~each sub-controller can only access system
information within its underlying region. To achieve these goals, we
start from considering the entire network as one region and design a
centralized controller satisfying~(i)-(iii). First, we consider an
open-loop finite-horizon optimal control problem, aiming to minimize
the overall accumulated control cost (to reflect requirement~(iii))
under two hard constraints corresponding to requirements~(i)-(ii). Due
to the non-convex and non-smooth nature of the optimization problem,
we then propose a convexification technique to obtain an sub-optimal
control input trajectory. To close the loop, for each state, its
control input is defined as the first step of the sub-optimal input
trajectory.  We show that the closed-loop system
satisfies~(i)-(ii). Finally, to achieve a distributed control
structure, we divide the network into several regions and separately
consider every region as a network.  The distributed controller for
each region is nothing but the centralized controller implemented on
it. By carefully taking into account the power flow interconnections
among the regions, we show that the closed-loop system also meets
requirements~(i)-(ii) under the distributed controller.
For reasons of space, all proofs are omitted and will
appear elsewhere.
%

\section{Preliminaries}\label{section:prelimiaries}
We introduce  notation and notions from graph theory.
               
\subsubsection{Notation}%
Let $\naturals$, $\real$, $\real_{>}$, and $\real_{\geqslant}$ denote
the set of natural, real, positive, and nonnegative real numbers,
respectively.  Variables are assumed to belong to the Euclidean space
if not specified otherwise. Denote $\ones_n$ and $\zeros_n$ in
$\real^n$ as the vector of all ones and zeros, respectively. For
$a\in\real$, denote $\lceil a \rceil$ as the biggest integer no bigger
than $a$. We let $\|\cdot\|$ denote the 2-norm on $\real^{n}$. For a
vector $b\in\real^{n}$, denote $b_{i}$ as its $i$th entry. For
$A\in\mathbb{R}^{m\times n}$, let $[A]_i$ and $[A]_{i,j}$ denote its
$i$th row and $(i,j)$th element. For any $c,d\in\naturals$, let
$[c,d]_{\naturals}= \left\{ x\in\naturals \big| c\leqslant x\leqslant
  d \right\}$.
        
\subsubsection{Algebraic graph theory}
We follow basic notions in algebraic graph theory
from~\cite{FB-JC-SM:08cor,NB:94}. An undirected graph is a pair
$\mathcal{G} = \mathcal(\mathcal{I},\mathcal{E})$, where $\mathcal{I}
= \{1,\dots,n\}$ is the vertex set and $\mathcal{E}=\{e_{1},\dots,
e_{m}\} \subseteq \mathcal{I} \times \mathcal{I}$ is the edge set.  A
path is an ordered sequence of vertices such that any pair of
consecutive vertices in the sequence is an edge of the graph. A graph
is connected if there exists a path between any two vertices. Two
nodes are neighbors if there exists an edge linking them. Denote
$\mathcal{N}(i)$ as the set of neighbors of node~$i$. For each edge
$e_{k} \in \mathcal{E}$ with vertices $i,j$, the orientation procedure
consists of choosing either $i$ or $j$ to be the positive end of
$e_{k}$ and the other vertex to be the negative end. 
The incidence
matrix $D=(d_{k i}) \in \mathbb{R}^{m \times n}$ associated with
$\mathcal{G}$ is defined as $ d_{k i} = 1$ if $i$ is the positive end
of $e_{k}$, $ d_{k i} = -1$ if $i$ is the negative end of $e_{k}$, and
$ d_{k i} = 0$ otherwise.
An induced subgraph
$\mathcal{G}_{\beta}=(\mathcal{I}_{\beta},\mathcal{E}_{\beta})$ of the
undirected graph $\mathcal{G} = \mathcal(\mathcal{I},\mathcal{E})$
satisfies $\mathcal{I}_{\beta}\subseteq\mathcal{I}$,
$\mathcal{E}_{\beta}\subseteq\mathcal{E}$, and
$(i,j)\in\mathcal{E}_{\beta}$ if $(i,j)\in\mathcal{E}$ with
$i,j\in\mathcal{I}_{\beta}$.  Additionally, for each
$\mathcal{G}_{\beta}$, let
$\mathcal{E}_{\beta}'\subseteq\mathcal{I}_{\beta}\times(\mathcal{I}\backslash\mathcal{I}_{\beta})$
be the collection of edges connecting $\mathcal{G}_{\beta}$ and the
rest of the network.

\section{Problem statement}\label{section:ps}
In this section we introduce the model for the power network dynamics
and state the control goals.
\subsection{Power network model}\label{subsection:model}
The power network is described by a connected undirected graph
$\mathcal{G}=(\mathcal{I},\mathcal{E})$, where
$\mathcal{I}=\{1,2,\cdots,n\}$ is the collection of buses and
$\mathcal{E}=\{e_{1},e_{2},\cdots,e_{m}\}\subseteq\mathcal{I}\times\mathcal{I}$
is the collection of transmission lines.  For each node
$i\in\mathcal{I}$, let $\omega_{i}\in\real$ and $p_{i}\in\real$ denote
the shifted voltage frequency relative to the nominal frequency, and
active power injection at node $i$, respectively.  Given an arbitrary
orientation on $\mathcal{G}$, for any edge with positive end $i$ and
negative end $j$, denote $f_{ij}$ as its signed power flow.  We
partition buses into $\II{u}$ and
$\mathcal{I}\backslash\II{u}$, depending on whether an external
control input is available to regulate the transient frequency
behavior for some buses.  The linearized power network
dynamics~\cite{NL-LC-CZ-SHL:14} described by states of power flows and
frequencies is
\begin{subequations}\label{eqn:dynamics-discrete}
  \begin{align}
    &\dot f_{ij}(t)=b_{ij}\left(\omega_{i}(t)-\omega_{j}(t)\right),\
    \forall (i,j)\in\mathcal{E}
    \\
    &M_{i}\dot\omega_{i}(t)=-E_{i}\omega_{i}(t)+q_{i}(t)+p_{i}(t)+u_{i}(t),
    \ \forall i\in\II{u},\label{eqn:dynamics-dis-2b}
    \\
    &M_{i}\dot\omega_{i}(t)=-E_{i}\omega_{i}(t)+q_{i}(t)+p_{i}(t), \
    \forall i\in\mathcal{I}\backslash\II{u},
    \\
    &q_{i}(t)\triangleq
\sum_{j:j\rightarrow i}f_{ji}(t)-\sum_{k:i\rightarrow k}f_{ik}(t),
  \end{align}
\end{subequations}%
where $b_{ij}\in\real_{>}$ is the susceptance of the line connecting
bus $i$ and~$j$, and $M_{i} \in \real_{\geqslant }$ and $E_{i} \in
\real_{\geqslant}$ are the inertia and damping coefficient of bus $i
\in \mathcal{I}$.  For simplicity, we assume that they are strictly
positive for every $i\in\mathcal{I}$. The term $q_{i}$ stands for the
aggregated electrical power injected to node $i$ from its neighboring
nodes, where $\left\{ j : j\rightarrow i \right\}$ is the shorthand
notation for $\left\{ j : j\in\mathcal{N}(i) \text{ and $j$ is the
    positive end of $(i,j)$} \right\}$.
        
For convenience, let $f\in\real^{m}$ and $\omega\in\real^{n}$ denote
the collection of $f_{ij}$ and $\omega_{i}$, respectively. Define
$p\in\real^{n}$ as the collection of all $p_{i}$'s. Let
$Y_{b}\in\real^{m\times m}$ be the diagonal matrix whose $k$th
diagonal item represents the susceptance of the transmission line
$e_{k}$ connecting bus $i$ and $j$, i.e., $[Y_{b}]_{k,k}=b_{ij},$ for
$ k=1,2,\cdots, m$.  Let $M \triangleq
\text{diag}(M_{1},M_{2},\cdots,M_{n})\in\real^{n\times n}$, $E
\triangleq \text{diag}(E_{1},E_{2},\cdots,E_{n})\in\real^{n\times n}$,
and $D\in\real^{m\times n}$ be the incidence matrix corresponding to
the given orientation procedure. We re-write
system~\eqref{eqn:dynamics-discrete} in compact form as
\begin{subequations}\label{sube:}\label{eqn:compact-form}
  \begin{align}
    \dot f(t)&=Y_{b}D\omega(t),
    \\
    M\dot\omega(t)&=-E\omega(t)-D^{T}f(t)+p(t)+u(t),\label{eqn:compact-form-2}
    \\
     & \hspace{-0.9cm}u(t) \!\in\! \mathbb{U} \! \triangleq \! \left\{ u\in\real^{n}   \big|
    \ \forall w\in[1,n]_{\naturals}, [u]_{w}=\left\{ \hspace{-.5cm}\begin{array}{ccc}
    & u_{w} & \text{if $w\in\II{u}$}     \notag
    \\
    & 0 & \text{otherwise}
    \end{array} \right.  \hspace*{-1ex}\right \} \! .
  \end{align}
\end{subequations}
For convenience, we use $ x\triangleq (f,\omega)\in\real^{m+n}$ to
denote the collection of all states.
We consider power injections that satisfy the following assumption.

\begin{assumption}\longthmtitle{Finite-time convergence of active
    power injection}\label{assumption:finite-convergence}
  For each $i\in\mathcal{I}$, $p_{i}$ is piecewise continuous and
  becomes a constant (denoted by $p_{i}^{*}$) after a finite time,
  i.e., there exists $0\leqslant \bar t<\infty$ such that
  $p_{i}(t)=p_{i}^{*}$ for every $i\in\mathcal{I}$ and every
  $t\geqslant \bar t$.
\end{assumption}

This type of power injections generalizes the common constant
injection assumption considered in the literature,
e.g.~\cite{CZ-UT-NL-SL:14,ARB-VV:00}.  Under
Assumption~\ref{assumption:finite-convergence}, one can show
that~\cite{YZ-JC:18-cdc1} for system~\eqref{eqn:compact-form} with
$u\equiv\zeros_{n}$, $(f(t),\omega(t))$ globally converges to the
unique equilibrium point $(f_{\infty},\omega_{\infty}\ones_{n})$
determined by power injection and network parameters, where
$\omega_{\infty}$ is called synchronized frequency.
        
\subsection{Control goal}
Our goal is to design distributed state-feedback control strategy for each
bus $i\in\II{u}$ that ensure that the frequency of buses of a
given targeted subset of $\II{u}$ stays within a safe
bound. Furthermore, we require that the designed controller
preserve the stability of the whole system. The above requirements are
explicitly formalized as follows:
        
\subsubsection{Frequency invariance requirement}
Given $\II{\omega}\subseteq\II{u}$, for each $i\in\II{\omega}$, let
$\underline\omega_{i}\in\real$ and $\bar\omega_{i}\in\real$ be such
that $\underline\omega_{i}<\bar\omega_{i}$.  We require that
$\omega_{i}(t)$ stays inside the safe region
$[\underline\omega_{i},\bar\omega_{i}]$ for any time $t>0$, provided
that the initial frequency $\omega_{i}(0)$ lies inside
$[\underline\omega_{i},\bar\omega_{i}]$.
        
\subsubsection{Asymptotic stability requirement}
Since the open-loop system state globally converges to
$(f_{\infty},\omega_{\infty}\ones_{n})$, we require that our
controller only affects the system's transients such that the
closed-loop also converges to the same equilibrium.

\subsubsection{Economic coordination requirement}
The controller $u$ should achieve the above two requirements by having
its sub-controller $u_{i}$ for $i\in\II{u}$ cooperate with
others to reduce the overall control effort.
        
\subsubsection{Distributed feedback realization}
Each sub-controller can only use state and power injection information
within a limited local region. This requirement makes the control
implementable for large-scale power networks, as each sub-controller
does not depend on global information.

Typically, the set $\II{\omega}$ consists of generator nodes with
over/underfrequency requirements, or nodes whose transient frequency
behaviors play a fundamental role in evaluating system
performance~\cite{NWM-KC-MS:11}. We have shown in~\cite{YZ-JC:18-auto}
that for every node $i\in\II{\omega}$, to guarantee its frequency
invariance, an external control signal has to be available at
node~$i$, i.e., $\II{\omega}$ must be a subset of $\II{u}$. Nodes in
$\II{u}\backslash\II{\omega}$, having no frequency requirement of
their own, assist nodes in $\II{\omega}$ to achieve frequency
invariance.

To meet the above four requirements, our strategy is to first
formulate an open-loop finite-horizon optimal control problem aiming
to minimize the overall control effort, and at the same time, having
two hard constraints reflecting frequency invariance and stability
requirements, respectively.  We then employ this formulation to design
a centralized feedback controller, and finally employ graph
decomposition to synthesize a distributed controller.
        
\section{Open-loop optimal control}
In this section, we formulate the optimization problem of interest.
Our goal is to minimize some cost function measuring control input
effort, subject to system dynamics, and frequency invariance and
asymptotic stability constraints. As the last constraints turn out to
be non-convex and non-smooth, we propose a convexification strategy by
generating a set of linear constraints. We build on this section later
to design the centralized and the distributed controllers.
        
\subsection{Open-loop finite-horizon optimal
  control}\label{subsection:option-loop}
We here first introduce a robust asymptotic stability condition with
respect to the open-loop equilibrium point.

\begin{lemma}\longthmtitle{Robust asymptotic stability
    condition} \label{prop:robust-as}
  For system~\eqref{eqn:compact-form}, suppose that the solution
  exists and is unique.  For every $i\in\II{u}$, let
  $\bar\omega_{i}^{\text{thr}}\in\real_{>}$ and $
  \underline\omega_{i}^{\text{thr}}\in\real_{<}$ be threshold values
  satisfying
  $\underline\omega_{i}^{\text{thr}}<\omega_{\infty}<\bar\omega_{i}^{\text{thr}}$. If
  for every $t\in\real_{\geqslant}$,
  \begin{subequations}\label{ineq:robust-stabilize-constraints}
    \begin{align}
      \omega_{i}(t)u_{i}(x(t),p(t))&\leqslant 0, \ \text{if
      }\omega_{i}(t)\nin
      (\underline\omega_{i}^{\text{thr}},\bar\omega_{i}^{\text{thr}}),\label{ineq:robust-stabilize-constraints-2a}
      \\
      u_{i}(x(t),p(t))&=0, \ \text{if }\omega_{i}(t)\in
      (\underline\omega_{i}^{\text{thr}},\bar\omega_{i}^{\text{thr}}),\label{ineq:robust-stabilize-constraints-2b}
    \end{align}
  \end{subequations}
  then under Assumption~\ref{assumption:finite-convergence}, $(f(t),\omega(t))\rightarrow(f_{\infty},\omega_{\infty})$ as
  $t\rightarrow\infty$. Furthermore, if $p(t)$ is time-invariant, then
  the closed-loop system is globally asymptotically stable.
\end{lemma}



Notice that the dependence of the robust asymptotic stability
condition~\eqref{ineq:robust-stabilize-constraints} on the equilibrium
point $(f_{\infty},\omega_{\infty}\ones_{n})$ is weak, as it requires
neither any information regarding $f_{\infty}$, nor a priori knowledge
of the exact value of $\omega_{\infty}$.  This reflects a practical
consideration under which the controller should still ensure
convergence and stability: although ideally $\omega_{\infty}$ is 0
when the load and supply are perfectly balanced (i.e.,
$\sum_{i=1}^{n}p_{i}^{*}=0$), due to imperfect estimation on the load
side and transmission loss, $\omega_{\infty}$ tends to slightly
deviate from~$0$.

With the stability condition being set, we now are ready to formally
introduce the finite-horizon optimal control problem, in which we aim
to minimize some cost function of control effort over a finite time
while respecting system dynamics, robust stability
condition~\eqref{ineq:robust-stabilize-constraints}, as well as the
frequency invariance condition.  As the power injection $p$ may not be
precisely known a priori, instead, for every $t\in\real_{\geqslant}$,
let a piecewise continuous signal $p^{fcst}_{t}:[t,t+\tilde
t]\rightarrow\real^{n}$ be its forecasted value for the first $\tilde
t$ seconds, starting from $t$. 
We adopt the following assumption on the forecast.

\begin{assumption}\longthmtitle{Forecast reveals current true value at
    frequency-controlled nodes}\label{assumption:forecast-injection}
  For every $i\in\II{\omega}$ and every $t\in\real_{\geqslant }$,
  assume that $p^{fcst}_{t,i}(t)=p_{i}(t)$, where $p^{fcst}_{t,i}(t)$
  is the $i$th component of $p^{fcst}_{t}(t)$.
\end{assumption}

A simple way to meet Assumption~\ref{assumption:forecast-injection} is
to, at every $i\in\II{\omega}$, first measure the power injection
$p_{i}(t)$ at the current time $t$, and then let
$p_{t,i}^{fcst}(\tau)=p_{i}(t)$ for every $\tau\in[t,t+\tilde t]$.
        
The optimization problem corresponding to the open-loop finite-horizon
optimal control is as follows,
\begin{subequations}\label{opti:nonlinear-continuous}
  \begin{alignat}{2}
    & & & \min_{f,\omega,u}\quad
    \sum_{i\in\II{u}}\int_{\tau_{0}}^{\tau_{0}+\tilde
      t}c_{i}u^{2}_{i}(\tau)\text{d}\tau\notag
    \\
    &\text{s.t.}&\quad &\dot f(\tau)=Y_{b}D\omega(\tau),\label{opti:nonlinear-a}
    \\
    &&&M\dot\omega(\tau)=-E\omega(t)-D^{T}f(\tau)+p^{fcst}_{\tau}(\tau)+u(\tau),\label{opti:nonlinear-b}
    \\
    &&&f(\tau_{0})=f_{0},\;
    \omega(\tau_{0})=\omega_{0},\label{opti:nonlinear-c}
    \\
    &&&u(\tau)\in\mathbb{U},\hspace{3.1cm}\;\forall
    \tau\in[\tau_{0},\tau_{0}+\tilde t],\label{opti:nonlinear-d}
    \\
    &&& \underline\omega_{i}\leqslant \omega_{i}(\tau)\leqslant
    \bar\omega_{i}, \hspace{0.9cm} \;\forall i\in\II{\omega},\;\forall
    \tau\in[\tau_{0},\tau_{0}+\tilde
    t],\label{ineq:hard-frequency-constraint}
    \\
    &&& (\omega,u)\in\Phi_{cont},\label{set:opti-nonlinear-continuous}
  \end{alignat}
\end{subequations}
where for every $i\in\II{u}$, $c_{i}\in\real_{>}$ corresponds to
the cost weight for $u_{i}$;
constraints~\eqref{opti:nonlinear-a}-\eqref{opti:nonlinear-c}
represent system dynamics and initial state;
constraint~\eqref{opti:nonlinear-d} reflects available controlled bus
indexes; constraint~\eqref{ineq:hard-frequency-constraint} refers to
the frequency invariance requirement, and
\begin{align*}
  \Phi_{cont}\triangleq\left\{ (\omega,u) \; \big| \;
    \eqref{ineq:robust-stabilize-constraints}\text{ holds }\forall t\in[\tau_{0}.\tau_{0}+\tilde t],\;\forall i\in \II{u}\right\}
\end{align*}
refers to the  stability condition established in Lemma~\ref{prop:robust-as}.

We refer to the optimization problem~\eqref{opti:nonlinear-continuous}
as $Q_{cont}(\mathcal{G}
,\II{u},\II{\omega},p_{\tau}^{fcst},f_{0},\omega_{0},\tau_{0})$ to
emphasize its dependence on the graph topology, controlled node
indexes, transient-frequency-constrained node indexes, power
injection, initial state, and initial time. If the context is clear,
we may just denote it as $Q_{cont}$ for brevity. We use the same
notational logic for optimization problems defined along the paper. In
addition, as we consider the hard frequency
constraint~\eqref{ineq:hard-frequency-constraint}, we assume
$(f_{0},\omega_{0})\in\Gamma$,~where
\begin{align}\label{set:initial-state}
  \Gamma\triangleq\left\{ (f,\omega) \big|\omega_{i}\leqslant
    \omega_{i}\leqslant \bar\omega_{i}, \hspace{0cm} \;\forall
    i\in\II{\omega} \right\},
\end{align}
so that the problem is well-defined.

        
In practice, a convenient way to approximate the functional solution
for $Q_{cont}$ is by discretization. Specially, here we discretize the
system periodically with time length $T\in\real_{>}$, and denote
$N\triangleq \lceil\tilde t/T\rceil$ as the total number of steps.
For every $k\in[0,N]_{\naturals}$, denote $\hat f (k),\hat\omega(k)$,
$\hat u(k)$, $\hat p^{fcst}(k)$ as the approximation of
$f(\tau_{0}+kT),\omega(\tau_{0}+kT)$, $u(\tau_{0}+kT)$ and
$p^{fcst}_{\tau}(\tau_{0}+kT)$, respectively, and let
\begin{subequations}\label{sube:eqn:traj}
  \begin{align}
    \hat F&\triangleq[\hat f(0),\hat f(1),\cdots,f(N)],
    \\
    \hat
    \Omega&\triangleq[\hat\omega(0),\hat\omega(1),\cdots,\hat\omega(N)],
    \\
    \hat P^{fcst}&\triangleq[\hat p^{fcst}(0),\hat
    p^{fcst}(1),\cdots,\hat p^{fcst}(N-1)],\label{eqn:p-G-graph}
    \\
    \hat U&\triangleq[\hat u(0),\hat u(1),\cdots,\hat u(N-1)],
  \end{align}
\end{subequations}
be the collection of power flow, frequency, predicted power injection,
and control input discrete trajectories, respectively. We formulate
the discrete version of $Q_{cont}$ as follows,
\begin{subequations}\label{opti:nonlinear}
  \begin{alignat}{2}
    & & & \min_{\hat F,\hat\Omega, \hat U}\quad g(\hat U)\triangleq
    \sum_{i\in\II{u}}\sum_{k=0}^{N-1}c_{i}\hat
    u^{2}_{i}(k)\notag
    \\
    &\text{s.t.}&\quad &\hat f(k+1)=\hat
    f(k)+TY_{b}D\hat\omega(k),\notag
    \\
    &&&M\hat\omega(k+1)=M\hat\omega(k)+T\big\{-E\hat\omega(k)-D^{T}\hat
    f(k)+\notag
    \\
    &&&\hspace{1.7cm}\hat p^{fcst}(k)+\hat u(k)\big\},\hspace{0cm}\;
    \forall k\in[0,N-1]_{\naturals},\label{opti:nonlinear-1}
    \\
    &&&\hat f(0)=f_{0},\
    \hat\omega(0)=\omega_{0},\label{opti:nonlinear-2}
    \\
    &&&\hat u(k)\in\mathbb{U}, \hspace{2.8cm}\; \forall
    k\in[0,N-1]_{\naturals},
    \\
    &&& \underline\omega_{i}\leqslant \hat\omega_{i}(k+1)\leqslant
    \bar\omega_{i}, \ \forall i\in\II{\omega},\ \forall
    k\in[0,N-1]_{\naturals},\hspace{-0.2cm}\label{opti:nonlinear-3}
    \\
    &&& (\hat\Omega,\hat U)\in\Phi_{disc},\label{set:opti-nonlinear}
  \end{alignat}
\end{subequations}
where 
\begin{align}
  \Phi_{disc}\triangleq\Big\{ (\hat\Omega,\hat U)\; \big| \; \forall
  i\in\II{u},\ \forall k\in[0,N-1]_{\naturals}, \text{ it holds
    that} \notag\\ &
  \hspace{-6.9cm}\hat\omega_{i}(k)\hat u_{i}(k)\leqslant 0, \ \text{if }\hat\omega_{i}(k)\nin (\underline\omega_{i}^{\text{thr}},\bar\omega_{i}^{\text{thr}}),\notag\\
  &\hspace{ -5.2cm} \hat u_{i}(k)=0, \ \text{if }\hat\omega_{i}(k)\in
  (\underline\omega_{i}^{\text{thr}},\bar\omega_{i}^{\text{thr}})\Big\}.
\end{align}
We refer to~\eqref{opti:nonlinear} as
$Q_{disc}(\mathcal{G},\II{u},\II{\omega},\hat
P^{fcst},f_{0},\omega_{0},\tau_{0})$.

\subsection{Constraint convexification}
From constraint~\eqref{set:opti-nonlinear}, one can see that the major
problem solving $Q_{disc}$ is to deal with the nonlinear and
non-smooth feasible set $\Phi_{disc}$.
To this end, we propose a convexification method that seeks to
identify a subset of $\Phi_{disc}$ consisting of only linear
constraints. This method relies on the notion of reference trajectory,
which is simply a trajectory $(\hat F,\hat\Omega,\hat U)$ that
satisfies~\eqref{opti:nonlinear}.
The following result formally states the convexification method using
a reference trajectory.

\begin{lemma}\longthmtitle{Convexification of nonlinear
    constraints} \label{lemma:convexificaiton}
  For any reference trajectory $(\hat
  F^{\text{ref}},\hat\Omega^{\text{ref}},\hat U^{\text{ref}})$, let
  \begin{align}
    \Phi_{cvx}\triangleq\Big\{ (\hat\Omega,\hat U)\; \big|  \; \forall i\in\II{u},\  \forall k\in[0,N-1]_{\naturals}, \text{ it holds that}   \notag\\
    &\hspace{-7.2cm}\hat\omega_{i}(k)\geqslant \bar\omega_{i}^{\text{thr}},\ \hat u_{i}(k)\leqslant 0, \ \text{if }\hat\omega_{i}^{\text{ref}}(k)\geqslant \bar\omega_{i}^{\text{thr}};\notag\\
    &\hspace{-7.2cm}\hat\omega_{i}(k)\leqslant  \underline\omega_{i}^{\text{thr}},\ \hat u_{i}(k)\geqslant  0, \ \text{if }\hat\omega_{i}^{\text{ref}}(k)\leqslant  \underline\omega_{i}^{\text{thr}};\notag\\
    &\hspace{ -6.5cm} \hat u_{i}(k)=0, \ \text{if
    }\underline\omega_{i}^{\text{thr}} <
    \hat\omega_{i}^{\text{ref}}(k)< \bar\omega_{i}^{\text{thr}}\Big\}.
  \end{align}
  Then, $\Phi_{cvx}$ is convex and non-empty, and
  $\Phi_{cvx}\subseteq\Phi_{disc}$.
\end{lemma} 
        
        
In light of Lemma~\ref{lemma:convexificaiton}, instead of directly
solving $Q_{disc}$ and given a reference trajectory, we alternatively
solve its convexified version by replacing $\Phi_{disc}$ by
$\Phi_{cvx}$ as follows,
\begin{subequations}\label{opti:linear}
  \begin{alignat}{2}
    & & & \min_{\hat F,\hat\Omega, \hat U}\quad g(\hat U)\notag
    \\
    &\text{s.t.}&\quad
    &~\eqref{opti:nonlinear-1}-\eqref{opti:nonlinear-3}\text{ hold},
    \\
    &&& (\hat\Omega,\hat U)\in\Phi_{cvx}.\label{set:opti-linear-phi}
  \end{alignat}
\end{subequations}
We refer to~\eqref{opti:linear} as
$Q_{cvx}(\mathcal{G},\II{u},\II{\omega},\hat
P^{fcst},f_{0},\omega_{0},\tau_{0})$.

\subsection{Reference trajectory generation}
We see that the key problem of the convexification 
is to find a suitable reference trajectory  to
approximate $\Phi_{disc}$ characterized by nonlinear constraints by
$\Phi_{cvx}$ containing only linear constraints. Based on our
previous work~\cite{YZ-JC:18-cdc1}, next we construct a specific
reference trajectory.
        
\begin{proposition}\longthmtitle{Generate reference
    trajectory} \label{prop:ref-generation}
  For every $i\in\II{u}$ and every $k\in[0,N-1]_{\naturals}$,
  suppose
  $\underline\omega_{i}<\underline\omega_{i}^{\text{thr}}<\omega_{\infty}
  < \bar\omega_{i}^{\text{thr}}<\bar\omega_{i}$, and $\bar\gamma_{i},\
  \underline\gamma_{i}\in\real_{>}$.  Define $\hat u^{\text{ref}}_{i}$
  in~\eqref{eqn:ref-controller} and let $\hat u^{\text{ref}}$ be the
  collection of $\hat u_{i}^{\text{ref}}$ over $i$. Define $\hat
  U^{\text{ref}}\triangleq[\hat u^{\text{ref}}(0),\hat
  u^{\text{ref}}(1),\cdots,\hat u^{\text{ref}}(N-1)]$, and $(\hat
  F^{\text{ref}},\hat \Omega^{\text{ref}})$ be the sate trajectory
  uniquely determined by~\eqref{opti:nonlinear-1}
  and~\eqref{opti:nonlinear-2} using $\hat U^{\text{ref}}$ as input.
  If $\underline\omega_{i}\leqslant \hat\omega_{i}(0)\leqslant
  \bar\omega_{i}$ holds for every $i\in\II{\omega}$, then there
  exists $\bar T\in\real_{>}$ such that for any $0<T\leqslant \bar T$,
  $(\hat F^{\text{ref}},\hat\Omega^{\text{ref}},\hat U^{\text{ref}})$
  is a reference trajectory.
\end{proposition}

\begin{figure*}[htb]
  \begin{align}\label{eqn:ref-controller}
    \hat u^{\text{ref}}_{i}(k)&\triangleq\left\{ \begin{array}{ccc} &
        \min\{0,\frac{\bar\gamma_{i}(\bar\omega_{i}-\hat\omega_{i}^{\text{ref}}(k))}{\hat\omega_{i}^{\text{ref}}(k)-\bar\omega_{i}^{\text{thr}}}-v_{i}(k)\}
        & \text{if }\hat\omega_{i}^{\text{ref}}(k)\geqslant
        \bar\omega_{i}^{\text{thr}},
        \\
        & 0 & \hspace{1cm}\text{if }\underline\omega_{i}^{\text{thr}}<
        \hat\omega_{i}^{\text{ref}}(k)< \bar\omega_{i}^{\text{thr}},
        \\
        &
        \max\{0,\frac{\underline\gamma_{i}(\underline\omega_{i}-\hat\omega_{i}^{\text{ref}}(k))}{\underline\omega_{i}^{\text{thr}}-\hat\omega_{i}^{\text{ref}}(k)}-v_{i}(k)\}
        & \text{if }\hat\omega_{i}^{\text{ref}}(k)\leqslant
        \underline\omega_{i}^{\text{thr}},
      \end{array} \right. \;\hspace{1cm}\forall i\in\II{\omega},\;\forall k\in[0,N-1]_{\naturals},
    \\
    \hat u_{i}^{\text{ref}}(k)&\triangleq0,\;\hspace{2cm}\forall i\in\II{u}\backslash\II{\omega},\;\forall k\in[0,N-1]_{\naturals},\notag
    \\v_{i}(k)&\triangleq \sum_{j:j\rightarrow i}\hat f_{ji}^{\text{ref}}(k)-\sum_{l:i\rightarrow l}\hat f_{il}^{\text{ref}}(k)+\hat p^{fcst}_{i}(k)-E_{i}\hat\omega_{i}^{\text{ref}}(k),\;\hspace{1cm}\forall i\in\II{\omega},\;\forall k\in[0,N-1]_{\naturals}.\notag
  \end{align}
  \hrulefill
\end{figure*}

From here on, we employ the specific reference trajectory
defined in Proposition~\ref{prop:ref-generation} in the convexification
method.  Notice that a small sampling length $T$ reduces the
discretization gap between $Q_{cont}$ and $Q_{disc}$, as well as
guarantees the qualification of $(\hat
F^{\text{ref}},\hat\Omega^{\text{ref}},\hat U^{\text{ref}})$ defined
in Proposition~\ref{prop:ref-generation} as a reference trajectory. On the
other hand, the number of constraints appearing in $Q_{cvx}$ grows
linearly with respect to $1/T$. Hence, it is of interest to understand
the trade-offs among the discretization accuracy, reference trajectory
qualification, and computational complexity.

\section{From centralized to distributed closed-loop receding horizon
  feedback}
        
%
%

In this section we close the loop on the system by defining the input
at a given state $(f(t),\omega(t))$ at time $t$ with a forecasted
power injection $p_{t}^{fcst}$ as the first step of the optimal
control input trajectory of
$Q_{cvx}(\mathcal{G},\II{u},\II{\omega},\hat
P^{fcst},f(t),\omega(t),t)$. We first consider a centralized control
strategy, where we assume that a single operator gathers global state
information, computes the control law, and broadcasts it to
corresponding sub-controllers. Based on this, we then propose a
distributed control strategy.
        
\subsection{Centralized control with stability and frequency
  invariance constraints}\label{subsection:c-MPC}
        
Formally, at time $t$, our centralized controller on one hand measures
the current state $(f(t),\omega(t))$, and on the other, forecasts a
power injection profile $p^{fcst}_{t}(\tau)$ with $\tau\in[t,t+\tilde
t]$ as well as its corresponding discretization $\hat P^{fcst}$
(cf.~\eqref{eqn:p-G-graph}).
Let $(\hat F^{*}_{cvx},\hat\Omega^{*}_{cvx},\hat U^{*}_{cvx})$ be the
optimal solution of
$Q_{cvx}(\mathcal{G},\II{u},\II{\omega},\hat
P^{fcst},f(t),\omega(t),t)$. The centralized control law is then given
by
\begin{align}\label{eqn:control-input-mpc}
  u(x(t),p^{fcst}_{t})\triangleq \hat u^{*}_{cvx}(0),
\end{align}
where $ u^{*}_{cvx}(0)$ is the first column of $\hat U^{*}_{cvx}$.
        
The following result states that the controller is able to guarantee
frequency invariance and, meanwhile, stabilize the system
without changing its open-loop equilibrium point.

\begin{theorem}\longthmtitle{Centralized control with stability and
    frequency invariance constraints}\label{thm:cen-control}
  Given power injection $p$ and any initial state
  $(f(0),\omega(0))\in\Gamma$, under
  Assumption~\ref{assumption:finite-convergence}
  and with sufficiently small sampling period $T$, the closed-loop
  system~\eqref{eqn:compact-form} with
  controller~\eqref{eqn:control-input-mpc} satisfies:
  \begin{enumerate}
  \item\label{item:convergence} $(f(t),\omega(t))\rightarrow
    (f_{\infty},\omega_{\infty}\ones_{n})$ as $t\rightarrow
    \infty$. Furthermore, if $p(t)$ is time-invariant, then the
    closed-loop system is asymptotically stable.
  \item\label{item:zero-input} For any $i\in\II{u}$ and any
    $t\in\real_{\geqslant }$, $u_{i}(x(t),p^{fcst}_{t})=0$ if
    $\omega_{i}(t)\in(\underline\omega_{i}^{\text{thr}},\bar\omega_{i}^{\text{thr}})$.
  \item\label{item:finite-convergence} 
     $u(x(t),p^{fcst}_{t})$ converges to $\zeros_{|\II{u}|}$ within a finite time.
  \item\label{item:frequency-invariance}Further under Assumption~\ref{assumption:forecast-injection},  for any $t\in\real_{\geqslant
    }$ and every $i\in\II{\omega}$, it holds $\omega_{i}(t)\in
    [\underline\omega_{i}, \bar\omega_{i}]$.
  \end{enumerate}
\end{theorem}

Note that to compute the centralized control signal defined
in~\eqref{eqn:control-input-mpc}, the operator should complete the
following procedures at every time: a)~collect state information and
forecast power injection of the entire network, b)~determine the
optimal trajectory $\hat U_{cvx}^{*}$ by solving $Q_{cvx}$, and
c)~broadcast the control signals to the corresponding
sub-controllers. Since the time to complete any of these three steps
grows with respect to the size of the network, it is impractical to
implement it for large-scale power networks.

\subsection{Distributed control using regional information}
Here we describe our approach to design a distributed control strategy
that retains the advantages of sub-controller cooperation with
stability and frequency invariance constraints.  The idea is to divide
the network into smaller regions, and have each sub-controller make
decisions based on the state and power injection prediction
information within its region.

\begin{assumption}\longthmtitle{Controlled nodes in induced
    subgraphs}\label{assumption:subgraph-node}
  Let $\mathcal{G}_{\beta}=(\mathcal{I}_{\beta},\mathcal{E}_{\beta}),\
  \beta\in[1,d]_{\naturals}$ be induced subgraphs of
  $\mathcal{G}$. Suppose
  \begin{subequations}\label{sube:assu-subgraph}
    \begin{align}
      &\II{u}\subseteq\bigcup_{\beta=1}^{d}\mathcal{I}_{\beta},\label{sube:assu-subgraph-1}
      \\
     &\mathcal{I}_{\alpha}\bigcap\mathcal{I}_{\beta} \bigcap \II{u}=
  \emptyset,\ \forall \alpha,\beta\in [1,d]_{\naturals}
      \text{ with } \alpha\neq\beta.\label{sube:assu-subgraph-2}
    \end{align}
  \end{subequations}
\end{assumption}

The induced subgraphs represent the regions of the network (note that
each controlled node is contained in one and only one region).  Our
distributed control strategy is to implement the centralized control
for every induced subgraph $\mathcal{G}_{\beta}$, where for every
$(i,j)\in\mathcal{E}_{\beta}'$, i.e., line connecting
$\mathcal{G}_{\beta}$ and the rest of the network, we treat its power
flow $f_{ij}(\tau)$ as an external power injection whose forecasted
value is a constant equaling its current value $f_{ij}(t)$ for every
$\tau\in[t,t+\tilde t]$.  Formally, denote for every
$i\in\mathcal{I}_{\beta}$,
\begin{align}\label{eqn:pfcst-f}
\hspace{-0.5cm} p_{t,\beta,i}^{fcst,f}(\tau)\triangleq\sum_{\substack{j\rightarrow
      i\\(i,j)\in\mathcal{E}'_{\beta}}}f_{ij}(t)-\sum_{\substack{i\rightarrow
      j\\(i,j)\in\mathcal{E}'_{\beta}}}f_{ij}(t),\;\forall \tau\in [t,t+\tilde
  t],
\end{align}
as the forecasted (starting from the current time $t$) power flow from
transmission lines in $\mathcal{E}_{\beta}'$ injecting into node
$i$. Let $p^{fcst,f}_{t,\beta}:[t,t+\tilde
t]\rightarrow\real^{|\mathcal{I}_{\beta}|}$ be the collection of all
such $p^{f}_{t,\beta,i}$'s with $i\in\mathcal{I}_{\beta}$. Also, let
$p^{fcst}_{t,\beta}:[t,t+\tilde
t]\rightarrow\real^{|\mathcal{I}_{\beta}|}$ be the collection of all
$p^{fcst}_{t,i}$'s with $i\in\mathcal{I}_{\beta}$, and denote $
p_{t,\beta}^{fcst,o}\triangleq
p^{fcst,f}_{t,\beta}+p^{fcst}_{t,\beta}$ as the overall forecasted
power injection for $\mathcal{G}_{\beta}$. Denote $\hat
P^{fcst,o}_{\beta}$ as its discretization.  Define
$\II{u}_{\beta}\triangleq\II{u}\bigcap\mathcal{I}_{\beta}$
(resp. $\II{\omega}_{\beta}\triangleq\II{\omega}\bigcap\mathcal{I}_{\beta}$)
as the collection of nodes within $\mathcal{G}_{\beta}$ with available
sub-controllers (resp. with frequency constraints). Let
$(f_{\beta},\omega_{\beta})\in\real^{|\mathcal{I}_{\beta}|+|\mathcal{E}_{\beta}|}$
be the collection of state within~$\mathcal{G}_{\beta}$.
        
        %

        
We are now ready to define the distributed control law. Similar
to~\eqref{eqn:control-input-mpc}, let $(\hat
F^{*}_{cvx,\beta},\hat\Omega^{*}_{cvx,\beta},\hat U^{*}_{cvx,\beta})$
be the optimal solution of
$Q_{cvx}(\mathcal{G}_{\beta},\II{u}_{\beta},\mathfrak{G_{\beta}},\hat
P^{fcst,o}_{\beta},f_{\beta}(t),\omega_{\beta}(t),t)$. The control law
is then given by
\vspace{-0.1cm}
\begin{align}\label{eqn:control-input-mpc-dis}
  u_{i}(x(t),p^{fcst}_{t})\triangleq \hat
  u^{*}_{i,cvx,\beta}(0),\;\forall i\in\II{u}.
\end{align}
\vspace{-0.1cm} where $u^{*}_{i,cvx,\beta}(0)$ is the $i$th entry of
$u^{*}_{cvx,\beta}(0)$, which is the first column of $\hat
U^{*}_{cvx,\beta}$. Note that, for any given $i\in\II{u}$,
$u_{i}(x(t),p^{fcst}_{t})$ only requires state and forecasted power
injection information within the corresponding induced subgraph
$\mathcal{G}_{\beta}$ and $\mathcal{E}_{\beta}'$: there is no need for
neither communication between any two induced subgraphs nor a priori
knowledge on topology or parameters of any other induced subgraphs.
        
We next state the properties of the control strategy.

\begin{proposition}\longthmtitle{Distributed control with stability
    and frequency invariance constraints}\label{prop:dis-control}
  Given power injection $p$ and initial state
  $(f(0),\omega(0))\in\Gamma$, under
  Assumptions~\ref{assumption:finite-convergence}
  and~\ref{assumption:subgraph-node}, with sufficiently small sampling
  period $T$, the following holds for system~\eqref{eqn:compact-form}
  with controller~\eqref{eqn:control-input-mpc-dis}:
  \begin{enumerate}
  \item\label{item:convergence-dis} $(f(t),\omega(t))\rightarrow
    (f_{\infty},\omega_{\infty}\ones_{n})$ as $t\rightarrow
    \infty$. Furthermore, if $p(t)$ is time-invariant, then the
    closed-loop system is asymptotically stable.
  \item\label{item:zero-input-dis} For any $i\in\II{u}$ and any
    $t\in\real_{\geqslant}$, $u_{i}(x(t),p_{t}^{fcst})=0$ if
    $\omega_{i}(t)\in(\underline\omega_{i}^{\text{thr}},\bar\omega_{i}^{\text{thr}})$.
  \item\label{item:finite-convergence-dis} $u_(x(t),p^{fcst}_{t})$'s converges to $\zeros_{|\II{u}|}$ within a finite time.
  \item\label{item:frequency-invariance-dis} Further, under Assumption~\ref{assumption:forecast-injection}, for any
    $t\in\real_{\geqslant }$ and every $i\in\II{\omega}$, it holds
    $\omega_{i}(t)\in [\underline\omega_{i}, \bar\omega_{i}]$.
  \end{enumerate}
\end{proposition}

\begin{figure}[htb]
  \vspace{-0.5cm}
  \centering%
  \includegraphics[width=0.9\linewidth]{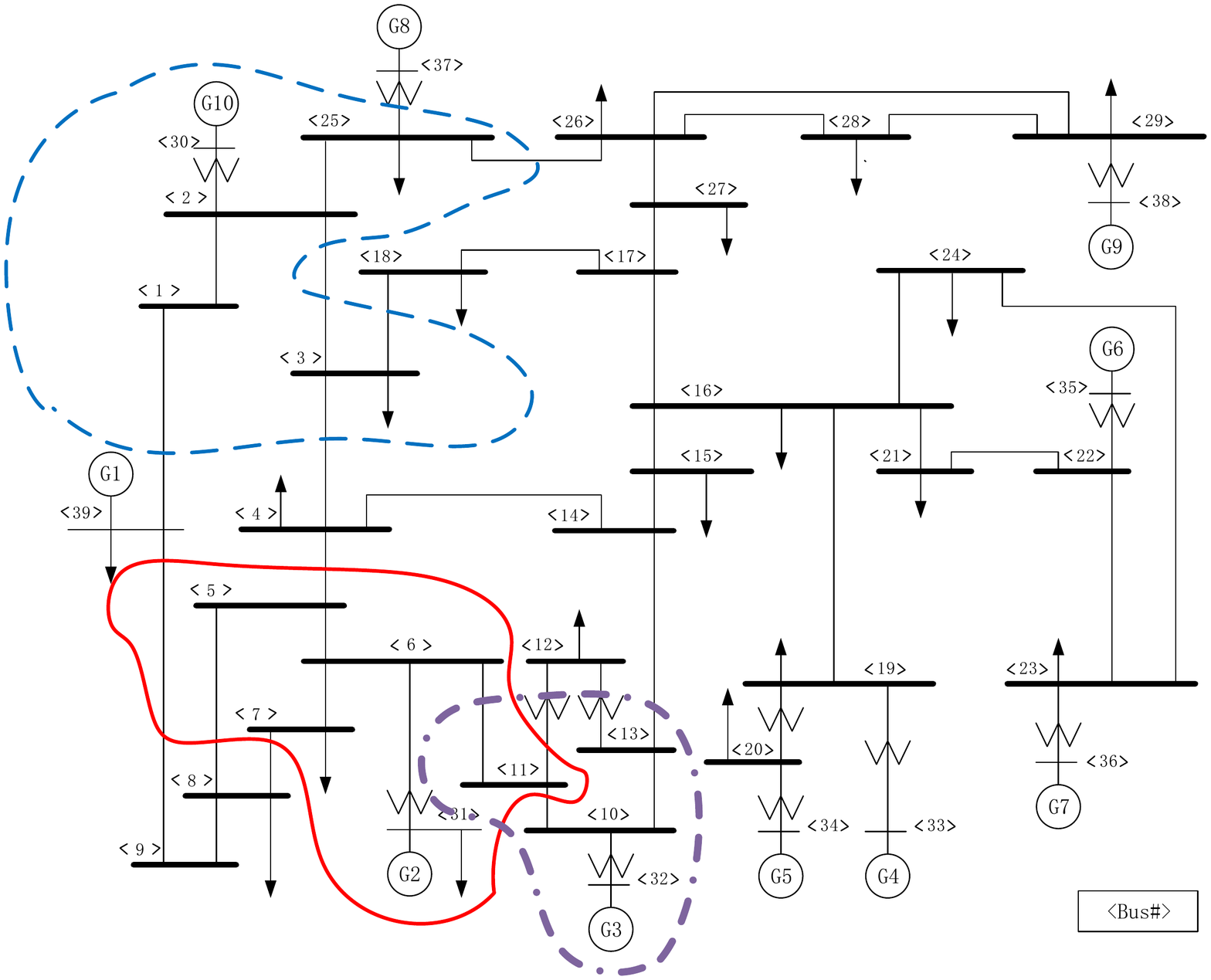}
  \vspace{-0.5cm}
  \caption{IEEE 39-bus power network.}\label{fig:IEEE39bus}
  \vspace*{-2ex}
\end{figure}

\section{Simulations}
        
We illustrate the performance of the distributed controller in the
IEEE 39-bus power network displayed in Fig.~\ref{fig:IEEE39bus}.
The network consists of 46 transmission lines and 10 generators,
serving a load of approximately 6GW.  We take the values of
susceptance $b_{ij}$ and rotational inertia $M_{i}$ for generator
nodes from the Power System Toolbox~\cite{KWC-JC-GR:09}. We also use
this toolbox to assign the initial power injection $p_{i}(0)$ for
every bus.  We assign all non-generator buses an uniform small inertia
$M_{i}=0.1$. Let the damping parameter be $D_{i}=1$ for all buses.
The initial state $( f(0),\omega(0))$ is chosen to be the equilibrium
with respect to the initial power injections. Let
$\II{\omega}=\{30,31,32\}$ be the three generators with transient
frequency requirements. We assign each of them a region containing its
2-hop neighbors. Let $\II{u}=\{3,7,25,30,31,32\}$ be the
collection of nodal indexes with sub-controllers.
\begin{figure}[tbh!]
  \centering
  \subfigure[\label{frequency-response-no-control-generator}]{\includegraphics[width=.32\linewidth]{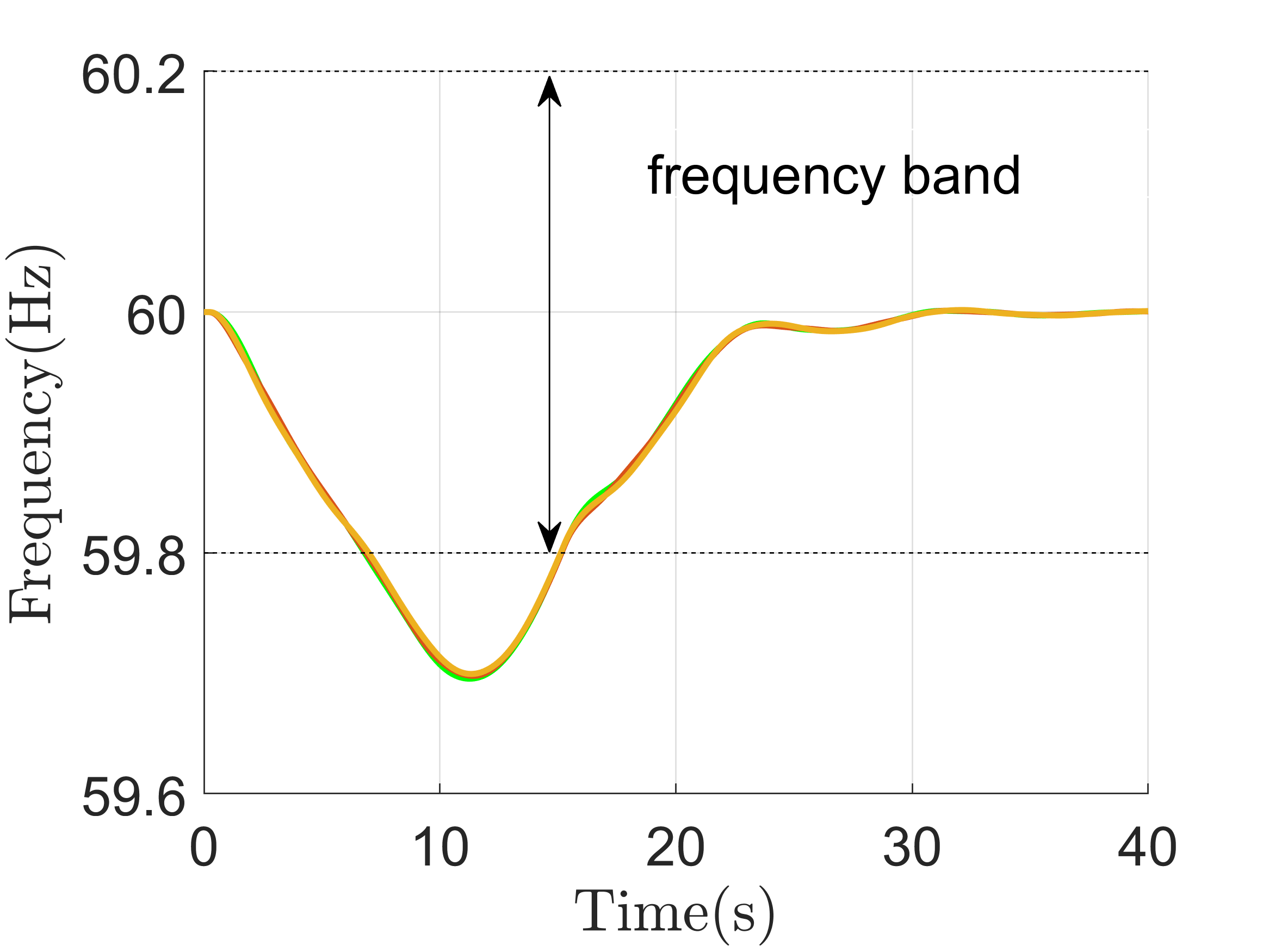}}%
  \subfigure[\label{frequency-response-with-control-generator}]{\includegraphics[width=.32\linewidth]{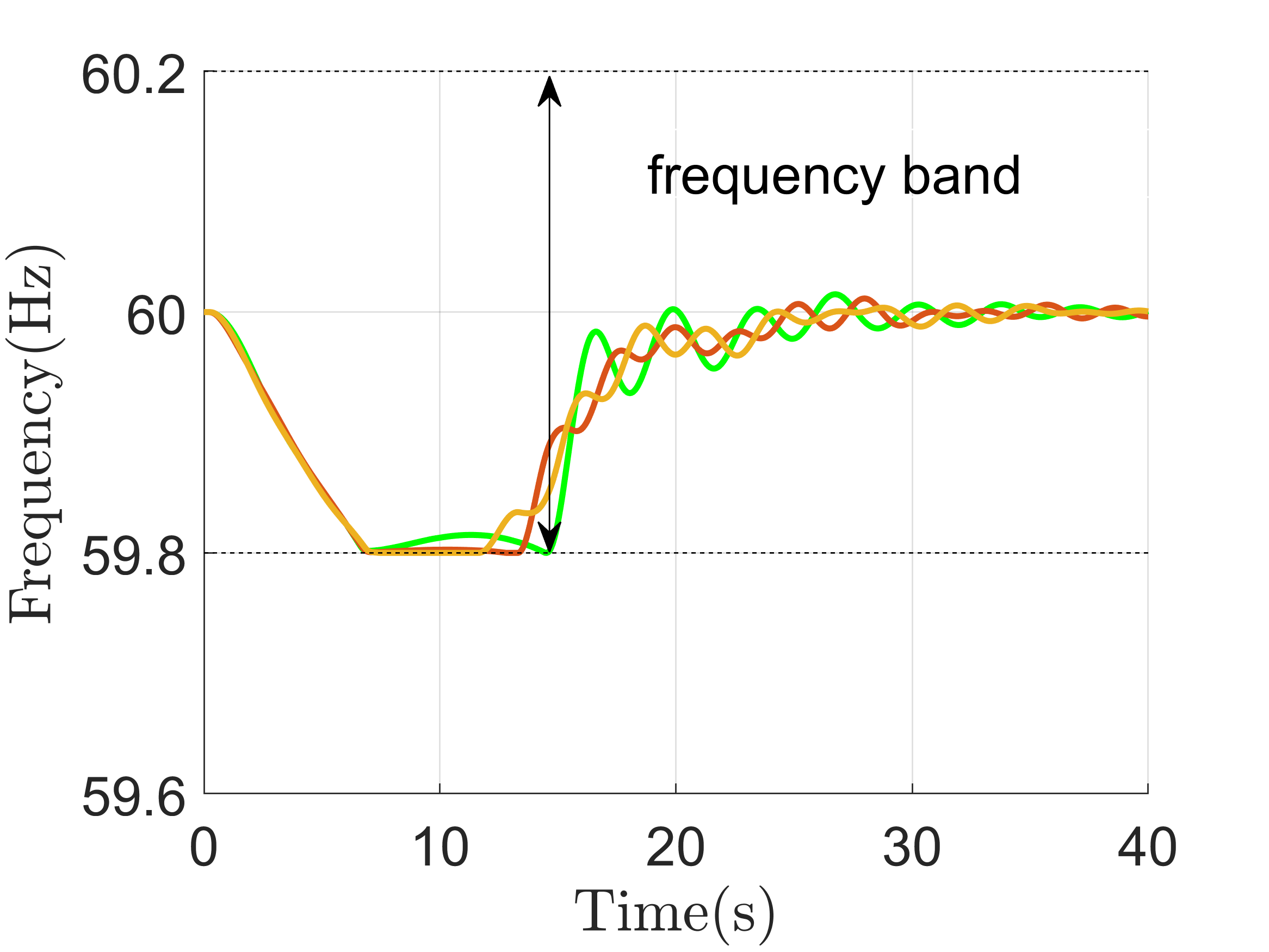}}%
  \subfigure[\label{frequency-input-mix-traj}]{\includegraphics[width=.32\linewidth]{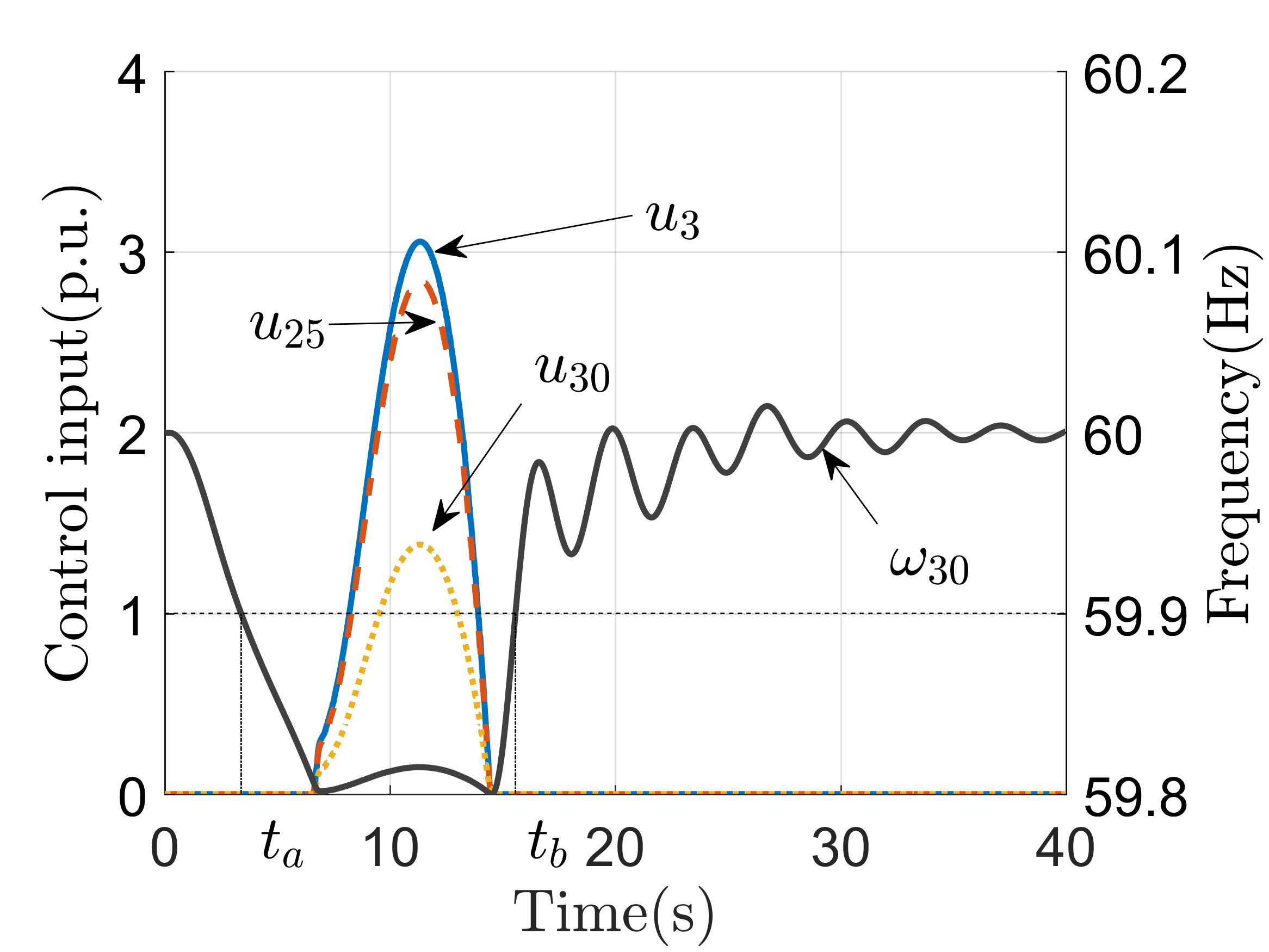}}
  \vspace{-0.3cm}
  \caption{Frequency and control input trajectories with and without
    distributed  frequency
    controller~\eqref{eqn:control-input-mpc-dis}.
    }\label{fig:trajectories}
\end{figure}
Notice that
Assumption~\ref{assumption:subgraph-node} holds in this scenario.  To
set up the optimization problem $Q_{cvx}$ so as to define our
controller~\eqref{eqn:control-input-mpc-dis}, for every
$i\in\II{u}$, we set $\bar\gamma_{i}=\underline\gamma_{i}=1$
required in~\eqref{eqn:ref-controller}, $c_{i}=2$ if
$i\in\II{\omega}$ and $c_{i}=1$ if
$i\in\II{u}\backslash\II{\omega}$, $T=0.001s$, $N=200$ so that
the predicted time horizon $\tilde t=0.2$s. Let
$\bar\omega_{i}=-\underline\omega_{i}=0.2$Hz and
$\bar\omega_{i}^{\text{thr}}=-\underline\omega_{i}^{\text{thr}}=0.1$Hz.
The nominal frequency is 60Hz, and hence the safe frequency region is
$[59.8\text{Hz},\ 60.2\text{Hz}]$. We assume
$p^{fcst}_{t}(\tau)=p(\tau)$ for every $\tau\in[t,t+\tilde t]$ for
simplicity.

        
We show that the proposed controller is able to maintain the targeted
generator frequencies within the safe region, provided that these
frequencies are initially in the safe region. We perturb all
non-generator nodes by a sinusoidal power injection whose magnitude is
proportional to the corresponding node's initial power
injection. Specifically, for every $i\in\{1,2,\cdots,29\}$, let
$p_{i}(t)=(1+\delta(t))p_{i}(0)$, where
\begin{align}\label{eqn:disturbance}
  \delta(t)=
  \begin{cases}
    0 & \hspace{-0.5cm}\text{if $t\leqslant0.5$ or $t\geqslant 15.5$,}
    \\
    0.3\sin(\pi\slash 15 (t-0.5))& \hspace{1.2cm}\text{otherwise.}
  \end{cases}
\end{align}
For $i\in\{30,31,\cdots,39\}$, let $p_{i}(t)\equiv p_{i}(0)$.
Fig.~\ref{fig:trajectories}\subref{frequency-response-no-control-generator}
shows the open-loop frequency responses of the 3 generators without
the controller, where all three trajectories exceed the lower bound
around 8s. As a comparison,
Fig.~\ref{fig:trajectories}\subref{frequency-response-with-control-generator}
shows the closed-loop response with the distributed controller, where
all frequencies stay within the safe bounds and converge to $60$Hz.
Fig.~\ref{fig:trajectories}\subref{frequency-input-mix-traj} shows
responses in the left-top region in Fig.~\ref{fig:IEEE39bus} (similar
results hold for the other two regions).  Notice that all three
control signals vanish within 20s. Since we assign a higher cost
weight on $u_{30}$, and the same weight on $u_{25}$ and $u_{3}$, the
latter two possess a similar trajectory, with magnitude higher than
the first one. On the other hand, notice that $u_{30}$ is always 0
while $\omega_{30}$ is above the lower frequency threshold denoted by
the dashed line. All these observations are consistent with
Proposition~\ref{prop:dis-control}.

\section{Conclusions}
We have proposed a centralized and a distributed frequency control on
power networks to maintain bus transient frequencies of interest
within given safe frequency intervals. We have shown that the
closed-loop system preserves the equilibrium point and convergence
propertues from the open-loop system, and the control input vanishes
in finite time. Furthermore, in the distributed control framework,
each sub-controller only requires regional information for feedback,
and sub-controllers within a same region cooperatively achieve
stability and frequency invariance by reducing the overall cost.
Future work will investigate the extension to nonlinear power flow
models andq the incorporation of optimization-based and real-time
control to reduce the computational time for controller implementation.

\bibliographystyle{IEEEtran}%
\bibliography{alias,JC,Main,Main-add}
        
\end{document}